# МОДЕЛЬ ТАБЛИЧНЫХ КОНСТРУКТОРСКИХ ДОКУМЕНТОВ ДЛЯ РАБОТЫ С ЭЛЕКТРОННЫМИ КАТАЛОГАМИ И СПЕЦИФИЦИРОВАНИЯ В САПР

В.В.Мигунов

Центр экономических и социальных исследований Республики Татарстан при Кабинете Министров Республики Татарстан

г.Казань, vmigunov@csp.kazan.ru

Моделирование табличных конструкторских документов (ТКД) в составе САПР необходимо для автоматизации специфицирования объектов, изображенных на чертежах, с использованием электронных каталогов. Модель разработана применительно к САПР реконструкции химических предприятий TechnoCAD GlassX, где результатом проектирования является чертеж с ТКД в виде спецификации, заказной спецификации, монтажной ведомости трубопровода, таблицы колодцев и др. Возможность адекватного моделирования ТКД предоставляется технологией совместного хранения в чертеже совокупности видимых геометрических элементов и невидимого параметрического представления ТКД, достаточного для генерации его изображения ("Табличный модуль"). Табличный модуль как элемент чертежа включает параметрическое представление и геометрическую часть, первичным является параметрическое представление.

ТКД представляется как массив записей, нулевая запись - шапка таблицы. Структура всех записей одинакова и задается последовательными актами деления блоков - прямоугольных фрагментов таблицы на части по вертикали и горизонтали, с признаками видимости этого деления в шапке и в области данных, на фиксированное или произвольное число частей, вплоть до неделимых блоков - клеток таблицы. Таким образом легко создаются ТКД с одинаковой шапкой, имеющие разделы (рубрики), как это предусматривает система проектной документации для строительства (СПДС), или не имеющие разделов. Видимые в шапке неделимые блоки имеют свойство - текст в шапке, который и помещается в чертеж и играет роль заголовков граф.

Каждому блоку может быть присвоен идентификатор объекта для поиска в электронных каталогах. Например, "Трубы". При указании на такой блок во время работы с табличным модулем предоставляется возможность выбора в электронных каталогах именно труб, то есть устанавливается автоматическая выборка в каталогах.

Неделимые блоки, видимые в области данных (клетки данных), имеют идентификацию свойств, значения которых помещаются в соответствующие графы. Например, свойство "Масса единицы., кг", "Наименование". Сведения из электронных каталогов также отвечают различным свойствам. Идентификация свойств по их номерам помещается в коде САПР и в файле правил генерации специфицирующих свойств по каталожным данным. После выбора в каталогах нужные свойства помещаются в соответствующие клетки ТКД. Задаются автоматически накладываемые ограничения на изделия в одном блоке (например, рабочие температура, давление, диаметр условного прохода). Значения ограничений берутся из заданной клетки и ускоряют выбор в каталогах за счет дополнительной выборки по условиям выполнения ограничений как вида изделий, так и их типоразмера (марки).

Для неделимых блоков предусмотрено также задание единиц измерения для их значений. Например, в таблицу заносятся пересчитанные значений давления в МПа, хотя при просмотре в каталогах могут встретиться и кгс/см2, и м вод.ст. Соответственно, и в электронных каталогах за полями таблиц закрепляются единицы измерения.

Автоматическая генерация спецификаций достигается путем хранения в элементах чертежа (модули аксонометрических схем, профилей наружных сетей водоснабжения и канализации, позиционных обозначений) невидимых специфицирующих свойств изделий с идентификацией свойств по номерам. Проектировщик задает область сбора информации (текущий чертеж или несколько записанных на диск, типы учитываемых модулей), и производится автоматическое заполнение нужных граф ТКД. Естественно, при этом отсутствует деление по разделам, которые затем создает проектировщик и выполняет автоматизированные операции фасовки строк, упорядочивания по задаваемой последовательности граф, сливания одинаковых и выделения общих частей наименований.

Иерархическая модель ТКД порождает сложности в программировании. Отсутствует понятие строки, и для идентификации выбираемой клетки требуется рекурсивный подсчет высот всех вышестоящих блоков данной записи. Однако эти трудности окупаются "интеллектом" ТКД. Например, при вставке в таблицу зависимости от точки указания вставляется именно тот прямоугольный блок, который является частью от деления на произвольное число частей по вертикали. При этом для фланцевых соединений, включающих сразу 5 изделий, вставляются одновременно по одной строке для фланца и прокладки и три строки для шпильки, гайки, шайбы (рис.рис. 1, 2).

Рис.1. Указание места вставки блока для специфицирования фланцевого соединения

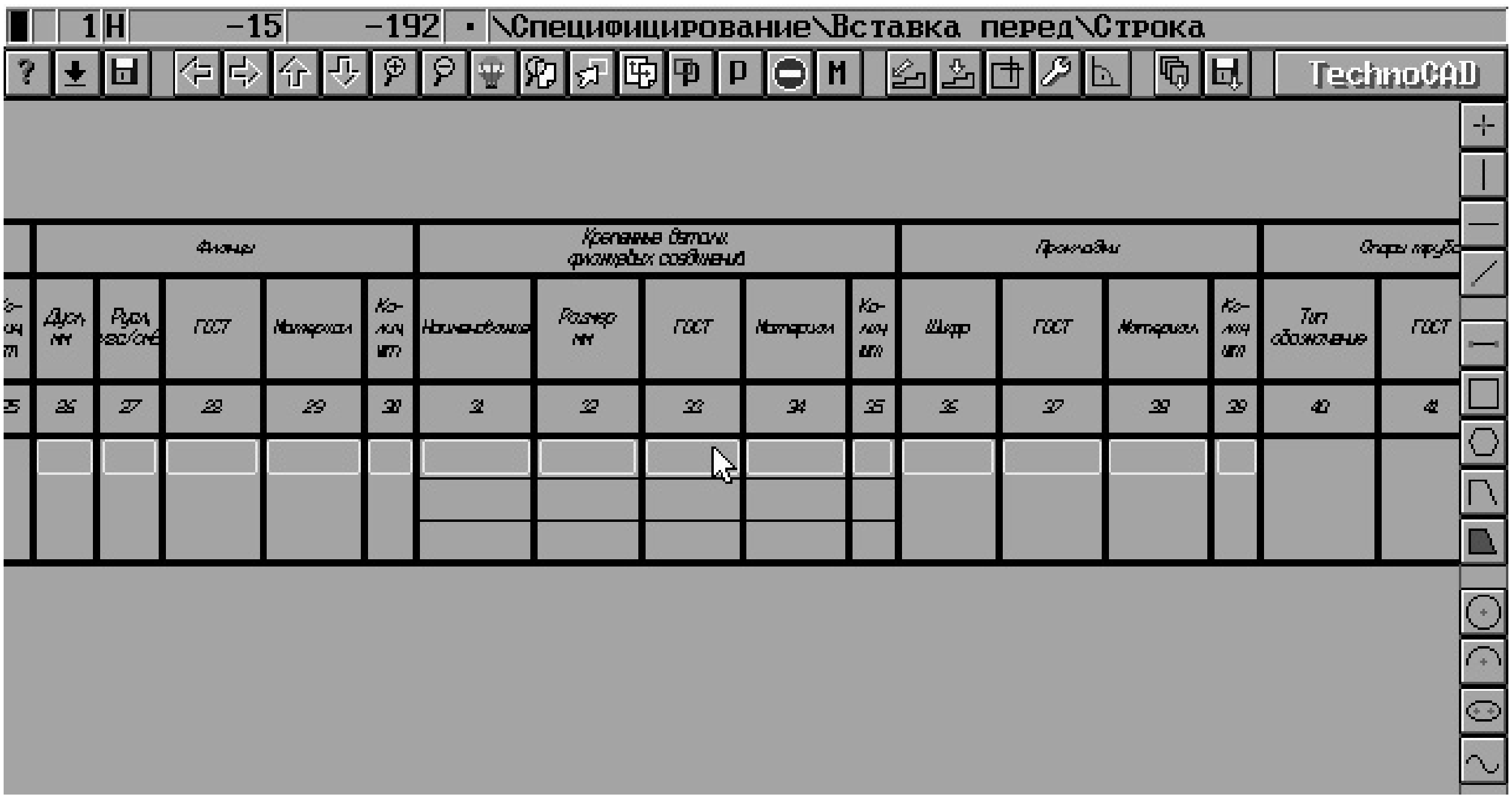

Рис.2. Результат вставки блока для специфицирования фланцевого соединения

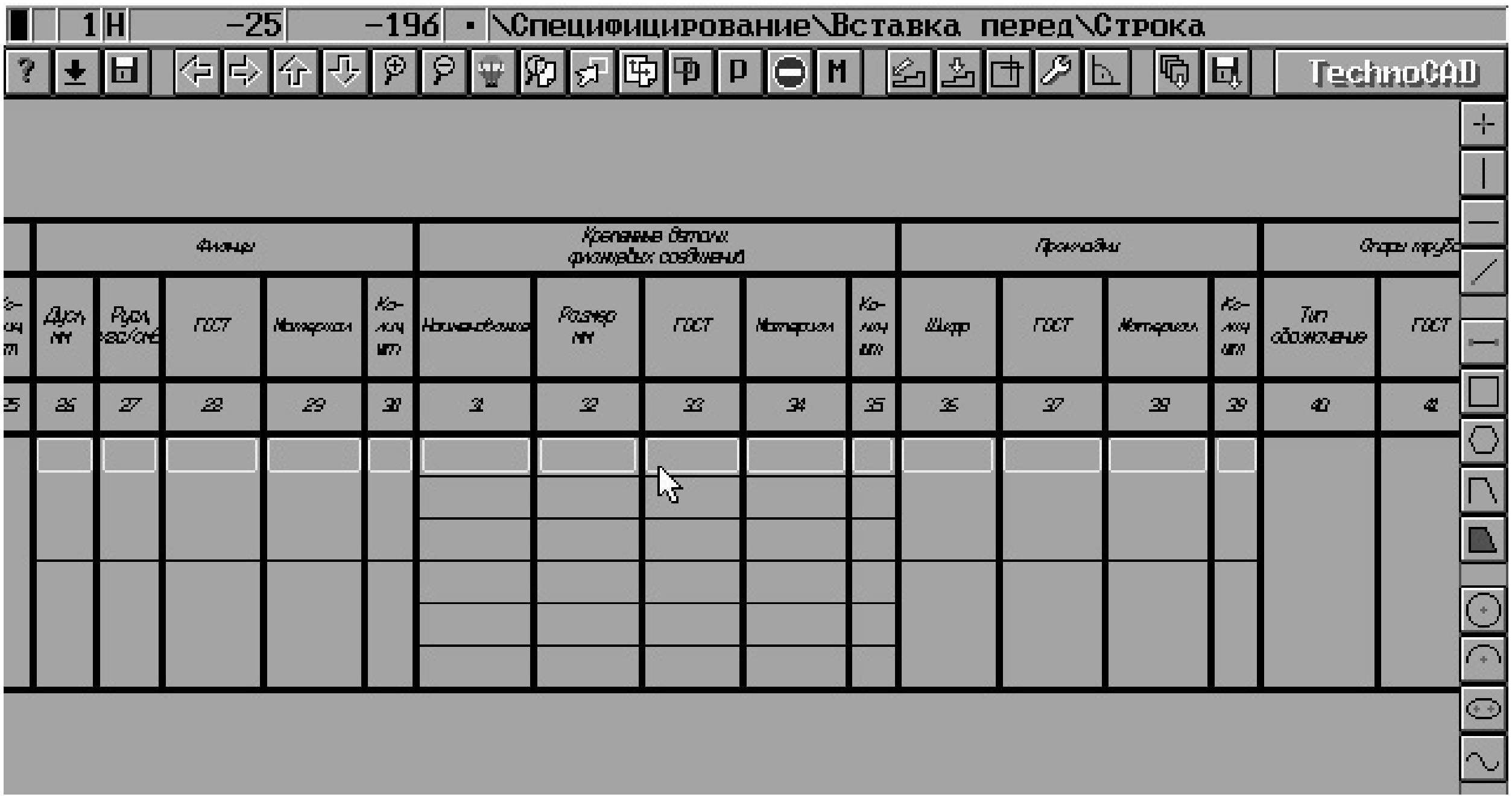

Рис.3. Помещение данных из экспликации в буфер изделий

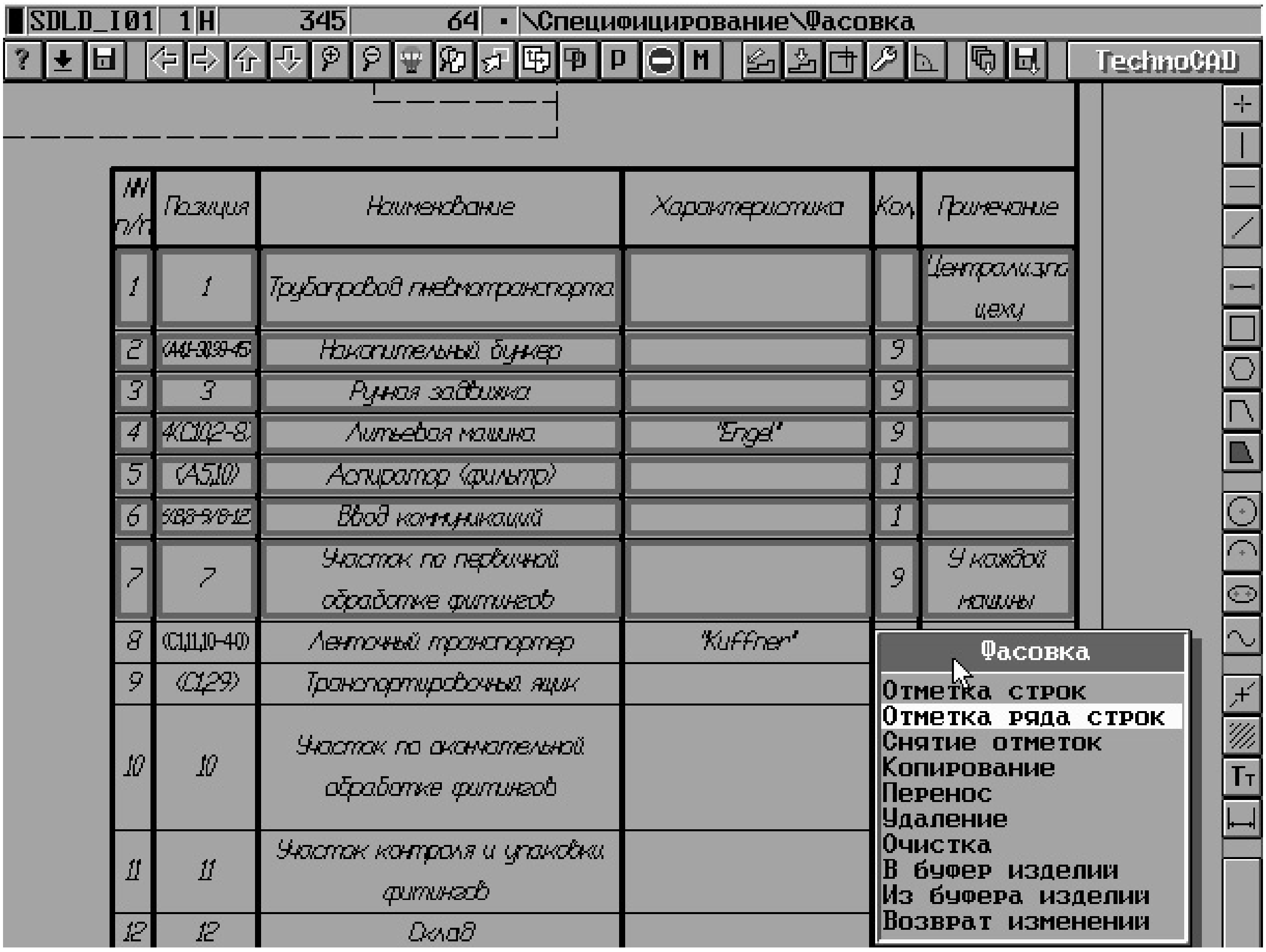

Для правки ТКД в специальном редакторе таблиц, где имеется доступ к электронным каталогам и не допускается иерархия блоков, из табличного модуля передается максимально возможная прямоугольная область, допускающая корректную работу с каталогами. Организован "буфер изделий", позволяющий переносить группу строк с одноименными специфицирующими свойствами между ТКД, в том числе различных видов (рис.рис.3,4). Эти преимущества достигаются не специализацией кода САПР для конкретных ТКД, а собственно структурой таблиц, задаваемой во внешнем файле.

Рис.4. Результат вставки из буфера изделий в спецификацию

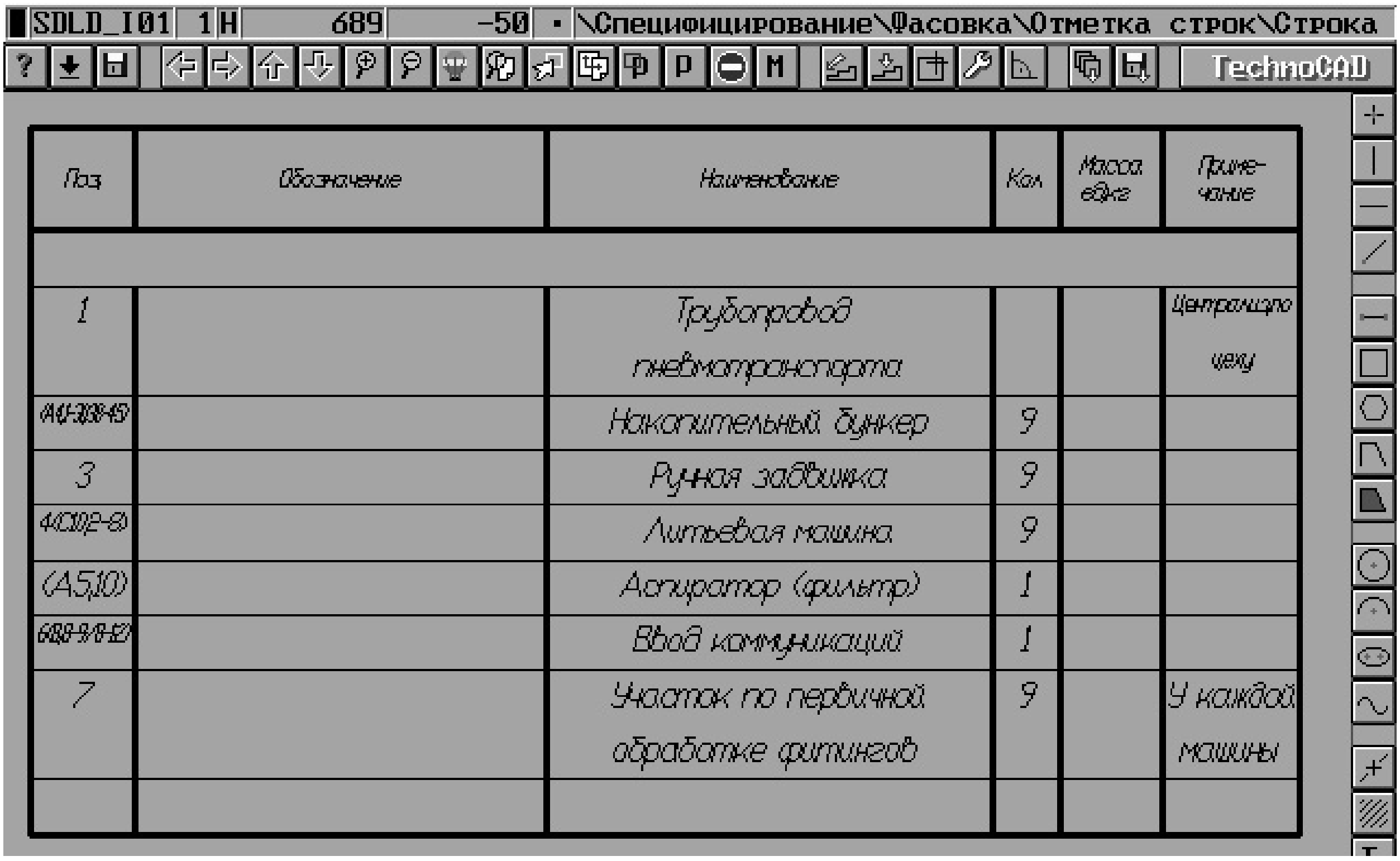

| Поз | Обозначение | Наименование | Кол | Масса ед.кг | Примечание |
|---|---|---|---|---|---|
| | | | | | |
| 1 | | Трубопровод пневмотранспорта | | | Централизованно цеху |
| [illegible] | | Накопительный бункер | 9 | | |
| 3 | | Ручная задвижка | 9 | | |
| [illegible] | | Литьевая машина | 9 | | |
| (А5,10) | | Аспиратор (фильтр) | 1 | | |
| [illegible] | | Ввод коммуникаций | 1 | | |
| 7 | | Участок по первичной обработке фитингов | 9 | | У каждой машины |
| | | | | | |

Сервис табличного модуля дает возможность продолжать ТКД влево или вправо кусками нужной высоты, иметь строки с номерами граф или повторять шапку, менять типы разделительных линий, шрифты текстов в каждой клетке и др. с мгновенной перегенерацией изображения. Комплект параметров табличного модуля можно записать на диск для последующей доработки и использования как прототипа.

Модель ТКД успешно эксплуатируется с 2000 года.

Подробнее о САПР TechnoCAD GlassX : www.technocad.narod.ru.